
\font\sixrm=cmr6
\font\sixi=cmmi6
\font\sixsy=cmsy6

\font\sevenrm=cmr7
\font\seveni=cmmi7
\font\sevensy=cmsy7

\font\twelverm=cmr12
\font\twelvei=cmmi12
\font\twelvesy=cmsy10 at 12pt
\font\twelveit=cmti12
\font\twelvesl=cmsl12
\font\twelvebf=cmbx12
\font\twelvett=cmtt12

\def\twelvepoint{%
\def\rm{\fam0\twelverm}%
\def\it{\fam\itfam\twelveit}%
\def\sl{\fam\slfam\twelvesl}%
\def\bf{\fam\bffam\twelvebf}%
\def\tt{\fam\ttfam\twelvett}%
\def\cal{\twelvesy}%
 \textfont0=\twelverm
  \scriptfont0=\sevenrm
  \scriptscriptfont0=\sixrm
 \textfont1=\twelvei
  \scriptfont1=\seveni
  \scriptscriptfont1=\sixi
 \textfont2=\twelvesy
  \scriptfont2=\sevensy
  \scriptscriptfont2=\sixsy
 \textfont3=\tenex
  \scriptfont3=\tenex
  \scriptscriptfont3=\tenex
 \textfont\itfam=\twelveit
 \textfont\slfam=\twelvesl
 \textfont\bffam=\twelvebf
 \textfont\ttfam=\twelvett
 \baselineskip=15pt
}

\font\sixteenrm=cmr17
\font\twentyrm=cmr17


\hsize     = 152mm
\vsize     = 215mm
\topskip   =  15pt
\parskip   =   0pt
\parindent =   0pt

\newskip\one
\one=15pt
\def\One{\vskip-\lastskip\vskip\one}

\newcount\LastMac
\def\Skipe{1}  
\def\Txe{2}    
\def\Hae{3}    
\def\Hbe{4}    

\def\SkipToFirstLine{
 \LastMac=\Skipe
 \dimen255=150pt
 \advance\dimen255 by -\pagetotal
 \vskip\dimen255
}

\def\Raggedright{%
 \rightskip=0pt plus \hsize
 \spaceskip=.3333em
 \xspaceskip=.5em
}

\def\Fullout{
 \rightskip=0pt
 \spaceskip=0pt
 \xspaceskip=0pt
}


\def\ct#1\par{
 \One
 \Raggedright
 \twentyrm\baselineskip=24pt
 #1
}

\def\ca#1\par{
 \One
 \Raggedright
 \sixteenrm\baselineskip=18pt
 \uppercase{#1}
}

\def\aa#1\par{
 \One
 \Raggedright
 \twelverm\baselineskip=15pt
 #1
}

\def\ha#1\par{
 \ifnum\LastMac=\Skipe \else \One\fi
 \LastMac=\Hae
 \Raggedright
 \twelvebf\baselineskip=15pt
 \uppercase{#1}
}

\def\hb#1\par{
 \LastMac=\Hbe
 \One
 \Raggedright
 \twelvebf\baselineskip=15pt
 #1
}

\def\tx{
 \ifnum\LastMac=\Hae \else
  \ifnum\LastMac=\Hbe \else
   \ifnum\LastMac=\Skipe \else \One
   \fi
  \fi
 \fi
 \LastMac=\Txe
 \Fullout
 \twelvepoint\rm
}

\output{\OutputPage}

\def\OutputPage{
 \shipout\vbox{\unvbox255}
}
\ct Non-Standard Phase Space Variables, Quantization, and Path Integrals,
or Little Ado about Much \par
\ca Michael P. Ryan, Jr.\par
\aa Instituto de Ciencias Nucleares

Universidad Nacional Aut\'onoma de M\'exico

A. Postal 70-546

M\'exico 04510 D.F.

MEXICO\par
\ca Sergio Hojman\par
\aa Departamento de F{\'\i}sica

Facultad de Ciencias

Universidad de Chile

Casilla 653

Santiago

CHILE\par

\ha 1. Introduction\par

\tx In this article we want to describe a long-term research project
that the authors plan to carry out over a period in the immediate
future.  We would like to outline the basic ideas of the project and
give a few preliminary calculations the bear on the validity of our
ideas as well as some speculations on where the research will lead.
We chose this volume to do this because it occured to us that each of
the components of our plan lies in a field that Charlie has touched
on at some point in his varied career.  These components are the
Hamiltonian formulation of the gravitational field equations, path
integral quantization, and quantum cosmology.  Such a wide-ranging
list shows how much we all owe to Charlie as a scientist and we hope
that our efforts will demonstrate our debt to his lead in these
fields.

\tx The genesis of our paper lies in a long-term concern that both of us
have had about the structure of quantum mechanics that has been the
subject of years of blackboard discussions between us (the
speculations of one of us [M.R.] go all the way back to
fondly-remembered discussions with Charlie and other Charlie-students
in Maryland).  The focal point of our ideas has always been the
cookbook nature of quantum mechanics as based on a Hamiltonian action
functional for the classical equations of motion and canonical
commutation relations derived from the Poisson bracket relations of
the variables in the action.  Perhaps the main among many objections
to this procedure are that not all classical equations of motion are
derivable from an action principle and the well-known fact that not
all canonical phase-space variables $(q_a, p_a)$ give a reasonable
quantum theory when simple quantization procedures are applied to
them.  In the quantum mechanics of mechanical systems and some fields
experiment has given enough guidance to be able to sidestep many of
these difficulties, but in the quantization of the gravitational
field there are no such experiments, and it is much more important to
have a well-formulated quantum theory that is as free from
ambiguities as possible.

\tx One of us (S.H.) has spent some time studying the Lagrangian and
Hamiltonian formulations of classical systems with an eye toward
quantization, and especially the quantization of the gravitational
field.  The other has studied quantum cosmologies as models for
quantum gravity.  We have already combined these studies in a paper
with Dario N\'u\~nez on the quantization of Bianchi-V cosmological
models using non-standard Hamiltonian techniques$^1$.  The project we
have in mind has several objectives and in this article we plan to
give examples from quantum cosmology of several of the possible lines
of investigation that we hope to pursue.

\tx Basically, we hope to give a better foundation to the idea of
quantization using non-standard Hamiltonian techniques by studying
the possible meaning of using very general phase-space variables as
a basis for quantization.  In the current paper we will not use the
Heisenberg picture (although this is not excluded as a future
possiblity), but will instead confine ourselves to Schr\"odinger and
path-integral methods.  One of the problems with Schr\"odinger
quantization in non-standard phase space variables is that some
operators are realized as derivatives, and it is vital to consider
the function space on which these derivatives operate.  As Bargmann$^2$
showed in the context of phase-space variables related to the usual
momentum and coordinate variables of the harmonic oscillator by a
complex ``canonical'' transformation, this function space need not be a
Hilbert space, but that one needs only demand that the space be
capable of being mapped onto a Hilbert space in a meaningful way.
While we are not in a position to develop a general mapping scheme
here, we will mention possible forms that such a mapping might take.

\tx With these goals in mind we will, in Sec. 2, sketch the problem of
quantization using non-standard phase-space variables, with some
emphasis on the difficulties of Schr\"odinger quantization.  In Sec. 3
we will use a Bianchi type-I cosmological model as a model for
quantum gravity and quantize it by means of BRST path integral
methods using a simple set of non-standard phase space variables.
Because of the speculative nature of the article we will not carry
any of our discussions too far, but only attempt enough to give the
broad outline of a picture we hope to fill in in the future.\par

\ha 2. Phase Space Variables in Quantum Mechanics.\par

\tx The usual formulation of quantum mechanics begins with an action
principle
$$ S = \int_{t_0}^{t_1} L(q_a, \dot q_a, t)dt = \int_{t_0}^{t_1}
[p_a \dot q_a - H(p_a, q_a )]dt, \eqno (2.1)$$
which in principle gives the classical equations of motion when one varies
$S$ with respect to $(q_a , \dot q_a)$ or $(p_a, q_a)$.  The standard
procedure is to convert the phase space variables $(q_a , p_a)$ into
operators $(\hat q_a , \hat p_a)$ and convert the Poisson bracket relations
$\{ q_a , q_b \} = 0 = \{p_a , p_b \}$, $ \{q_a, p_b \} = \delta_{ab}$
into operator commutator relations $[\hat q_a , \hat q_b ] = 0 =
[\hat p_a , \hat p_b ]$, $[\hat q_a , \hat p_b ] = i\delta_{ab}$.  One
also takes any function $O(q_a , p_b)$ and converts it to an operator
$\hat O (\hat q_a , \hat p_b)$ which has (modulo factor -ordering problems)
the same functional form as $O$.  The most important function for the
quantum dynamics of the system is the Hamiltonian operator $\hat H(
\hat q_a , \hat p_b)$.  In this article we will concern ourselves mainly
with the Schr\"odinger picture, where the momentum or coordinate
variables are realized as derivative operators on a function space
consisting of functions of eigenvalues of the other operators and the
time $t$.  Note that we say ``function space'' rather than Hilbert
space; this is deliberate.

\tx This procedure leads to a Schr\"odinger equation
$$ \hat H \Psi (q_a , t) = i {{\partial \Psi}\over {\partial t}} (q_a ,t)
\eqno (2.2)$$
[or the equivalent for a momentum space wave function, $\Phi (p_a , t)$].
If the solutions to (2.2) form a Hilbert space, we can define a positive-
definite probability density $\rho (q_a , t) = \Psi^* \Psi (q_a, t)$
on the space of solutions.

\tx This simple-minded precis of quantum theory ignores a large number of
well-known problems, some of which have at least tentative solutions, and
we will assume the reader is familiar with these.  The quantization of
constrained systems, the quantum theory of relativistic systems, and
factor ordering difficulties are among them.  The problems we plan to discuss
in most detail are basically difficulties associated with the process
of Schr\"odinger quantization we outlined above.  It is well known (and
usually cheerfully ignored) that the process is only valid for equations
of motion that admit an action principle of the form (2.1), and the complete
procedure requires a Hamiltonian form of the action, and that the phase
space formulation where one realizes operators as derivatives is only
supposed to be valid for a restricted class of phase space coordinates.

\tx One of the best known examples of the failure of straightforward
Schr\"odinger quantization is the motion of a particle in a central
potential.  In Cartesian coordinates we have
$$S = \int [p_x \dot x + p_y \dot y + p_z \dot z - \{{{1}\over {2}}
(p_x^2 + p_y^2 + p_z^2 ) + V(x^2 + y^2 + z^2 ) \} ]dt \eqno (2.3)$$
and if we realize $\hat p_x$, $\hat p_y$, $\hat p_z$ as $-i\partial_x$,
$-i\partial_y$, $-i \partial _z$, then we arrive at a correct Schr\"odinger
equation, $i \partial \Psi / \partial t = \hat H \Psi$.  If we make the
canonical transformation to spherical coordinates, $x,y,z$ $\rightarrow$
$r,\theta, \phi$,\hskip 1em $p_x, p_y, p_z$ $\rightarrow$ $p_r, p_{\theta},
p_{\phi}$,
the action becomes
$$S = \int [p_r \dot r + p_{\theta} \dot \theta + p_{\phi} \dot \phi -
\{ {{1}\over {2}}(p_r^2 + {{p_{\theta} ^2}\over {r^2}} + {{p_{\phi}^2}\over
{r^2 \sin^2 \theta}}) + V(r) \} ]dt, \eqno (2.4)$$
and if we try to realize the momentum operators $\hat p_r, \hat p_{\theta},
\hat p_{\phi}$ as $-i\partial_r, -i\partial_{\theta}, -i\partial_{\phi}$,
the resulting Schr\"odinger equation is ``incorrect'',
at least given a naive interpretation of the resulting wave function
$\Psi (r, \theta , \phi , t)$.  Since it can be shown that any infinitesimal
canonical transformation is equivalent to a unitary transformation acting
on the corresponding operators, is is only transformations such as those
to spherical coordinates that cannot be built up from infinitesimal
transformations that will cause such problems.

\tx In this article we would like to speculate on possible solutions to
these problems.  We have discussed problems that can be broken down into
three main points: 1)How can one quantize a system described by any
phase space variables? 2) How can one quantize systems whose equations of
motion do not come from an action principle?; and 3)Can one quantize a
system in the Schr\"odinger representation in non-standard phase-space
variables?  In order to make at least some steps toward answering these
questions, we will begin with an outline of the symplectic approach to the
canonical formulation of the equations of motion.

\tx In principle one needs a first order formulation of the equations of
motion of a system described by a set of variables $x^a$, $a = 1,\cdots, 2n$.
It is usual to suppose that half of these variables are coordinates and
half are velocities or momenta.  The equations of motion are
$$\dot x^a = f^a (x^b). \eqno (2.5)$$
A canonical set of equations that reproduces these is
$$ \{x^a , x^b \} = J^{ab}; \eqno (2.6a)$$
$$ \dot x^a = \{x^a , H \} = J^{ab} {{\partial H}\over {\partial x^b}} =
f^a, \eqno (2.6b)$$
where for consistency of the equations of motion $J^{ab}$ must obey the Jacobi
identity,
$${J^{ab}}_{,d}J^{dc} + {J^{bc}}_{,d}J^{da} + {J^{ca}}_{,d}
J^{db} =0. \eqno (2.7)$$
The symplectic structure matrix $J^{ab}$ is usually assumed to have the
form
$$  J^{ab} = \left [ \matrix {0&I_N\cr -I_N&0\cr} \right ] ,
    \eqno (2.8)$$
where $I_N$ is the $n \times n$ unit matrix,
so that half of the $x^a$ are momenta and the other half configuration
variables.  Of course, one must still be able to find a Hamiltonian $H$
that gives (2.6b).

\tx In an unpublished work reported by Dyson$^3$, Feynman recognized that the
form of $J^{ab}$ given by (2.8) is very restrictive and that relaxing
this requirement leads to easier solution of certain problems, but dropped
 the idea because he felt it was unphysical.  Recently Hojman and Shepley$^4$
have attempted to use Feynman's extension of symplectic theory to find
new methods of quantization.  They have shown that there exists a
Hamiltonian $H$ for any system of the form (2.6), and a $J^{ab}$ can be
constructed for such a system if one knows $2n$ constants of the motion
$C_i$, ($2n-1$ of which do not depend explicitly on time), that is knows
them explicitly as functions of the coordinates (a fairly strong requirement,
equivalent to knowing the full classical solution).  This $J^{ab}$ may be
constructed by summing elements of the basic form$^5$
$$J_1^{ab} = \lambda (x^c) \varepsilon ^{ab\mu_1 \cdots \mu_{2n-2}}
C_{1,\mu_1}
\cdots C_{2n-2,\mu_{2n-2}}, \eqno (2.9)$$
where $C_{,\mu} \equiv {{\partial C}\over {\partial x^{\mu}}}$, $\varepsilon
^{ab\mu_1 \cdots \mu_{2n-2}}$ is the $2n$-index Levi-Civita symbol, and
$\lambda(x^c)$ is a function of the phase space coordinates that will be
explained below.  This $J^{ab}$ satisfies the Jacobi identity.  The $C_1 \cdots
 C_{2n-1}$ are time independent constants of the motion. The Hamiltonian
is defined as $H = C_{2n-1}$, along with $C_{2n} = t + d_{2n}$, where
$d_{2n}$ is time independent.  This can always be achieved by a change
of coordinates.  It is easy to realize that $\lambda(x^c)$ may always be
chosen so that $J_1^{ab} {{\partial H}\over {\partial x^b}} = f^a$.
There is considerable freedom in this formalism in selecting the
Hamiltonian $H$.
The main advantage of this formulation is that it allows one to find
Hamiltonians for systems that do not admit an action principle (an
example that will be mentioned below is the set of Class B Bianchi models
when the isometries are imposed directly in the Hilbert action).  A
disadvantage (which we feel may not be a true disadvantage) is that the
Hamiltonian is not unique.

\tx Perhaps the best way of characterizing this concept is to imagine that
the form of $J^{ab}$ given in (2.8) is the phase space equivalent of a
``Minkowski metric'', and that phase space ``general coordinate
transformations'' as opposed to ordinary canonical transformations, the
equivalent of Lorentz transformations, will naturally change the metric
form while preserving the equations of motion (2.6).  Notice that one can
make very general phase space transformations, including transformations to
sets of variables that are all constants of the motion, the equivalent of
the procedure that leads to the Hamilton-Jacobi formulation of classical
mechanics.

\tx In the future we are planning to use these ideas to attempt to build up a
quantum theory and to apply it to a list of problems.  Perhaps one of the
most interesting possibilities will be its application to quantum gravity.
We have already made a first attempt, applying it to the quantization of
Bianchi Type V models in quantum cosmology.

\tx There are essentially three paths we could take.  One would be the use
of the Heisenberg picture, since the equations of motion (2.6) are written
in a form that is particularly amenable to this representation.  While
we plan to investigate this line later, our interest in quantum gravity
has, given the direction that field has taken lately, led us to consider
first the two other major methods of quantization, both of which are a
little more difficult in the context of Eqns. (2.6).  One is Schr\"odinger
quantization and the other is path integral quantization.  We will discuss
the difficulties of Schr\"odinger quantization below, while path integral
quantization will be the subject of Sec. 3. The symplectic structure
language that we use is reminiscent of that used in geometric
quantization$^6$, although the aims and methods are different, and
only canonical transformations (``Lorentz'') transformations are
allowed there.  It is possible that there may be some natural points
of contact between our ideas and that theory.

\tx In principle, Schr\"odinger quantization is not difficult, at least
mechanically.  We must select half of the operators $\hat x^a$ as
multiplication operators so the ``wave function'' will be a function
of the eigenvalues of these operators and, in principle, time.  The rest
of the operators will be realized as derivative operators with respect
to their ``conjugates''.  We can then construct a Schr\"odinger equation
$$ i {{\partial \Psi}\over {\partial t}} (x^a,t) = \hat H (x^a,
-i\partial_{x^a} ) \Psi (x^a,t). \eqno (2.10)$$
There are two main problems here, one obvious and the other a bit more
subtle.  One is that when $J^{ab}$ is no longer of the form (2.8) it
becomes a tricky problem to decide which variables are conjugate to which.
In the examples we have considered so far this problem can be handled,
but we do not yet have a general procedure.  The second problem is: To
what function space do the solutions $\Psi (x^a, t)$ belong?  There  is no
guarantee that they form a Hilbert space, and this is often considered
a fatal defect.  However, the original Bargmann formulation for the
harmonic oscillator$^2$, that is the germ of
the Ashtekar variables$^7$, addressed
exactly this question.  The complex ``canonical'' transformation
$\bar \Pi \equiv
{{1}\over {\sqrt {2}}} (p + iq)$, $ X \equiv {{1}\over {\sqrt{2}}}(q + ip)$
leads to an action (total derivatives dropped)
$$ S = \int [\bar \Pi \dot X + iX\bar \Pi] dt, \eqno (2.11)$$
which can be quantized in the Schr\"odinger representation by realizing
$\bar \Pi$ as $-i\partial / \partial X$.  This leads to a very simple
Schr\"odinger equation of the form
$$ \hat H \Psi (X,t) = -X{{\partial \Psi}\over {\partial X}} = i{{\partial
\Psi}\over {\partial t}}, \eqno (2.12)$$
first given by Fock$^8$.  What Bargmann showed
was that the function space upon
which $\partial / \partial X$ operates is not a Hilbert space, but
that there exists a kernel $K(q,X)$ which maps the space of functions
${\cal F} (X)$ of solutions to (2.12)
onto the usual Hilbert space ${\cal H} (q)$, that is
$$ \psi (q,t) = \int K(q,X) \Psi (X,t) dX, \eqno (2.13)$$
where $dX$ means $dpdq$.  This, of course, means that wave functions
$\Psi (q_a, p_a , t)$ need not be wave functions in the usual sense.
However, we have achieved simpler equations (note that [2.13] is much
simpler than the usual harmonic-oscillator Schr\"odinger equation) at the
cost of finding the kernel $K$ that maps the new wave functions into
useful wave functions that have the properties usually associated with them.
In the case of the harmonic oscillator, finding $K$ is relatively easy, but
for more general phase space variables it may become quite difficult.  We
plan to investigate this in the future.

\tx In a first attempt to apply some of our ideas to quantum gravity, the
authors and Dario N\'u\~nez quantized Bianchi type V cosmological models
as quantum cosmologies with one-sixth the logarithm of the determinant
of the three-metric as an internal time$^1$.  The Einstein equations in
this case do not result from the variation of the Hilbert action restricted
by symmetry, but we were able to find a Hamiltonian $H$ in the sense of
Eqns. (2.6) and use it for Schr\"odinger quantization of the theory.  We
used a simple-minded mapping of the resulting wave function to one that
could be used to construct a Hilbert space rather than attempt to construct
a Bargmann-like kernel.  We argued that given the confused state of the
current interpretation of the wave function of the universe, that our
solution has as much claim as any other to be a viable candidate for
such a wave function.

\tx We will not give the details here of the Bianchi V calculation, but will
retreat to a simpler model, the Bianchi type I cosmologies.  In the section
that follows we will go on to the next part of the program outlined in the
Introduction and apply BRST path-integral quantization to these models
in non-standard phase space coordinates.  We will again use a simple
mapping of the wave function we obtain rather than attempt to develop here
the more complicated integral mapping outlined above.

\ha 3. Path Integral Quantization of Bianchi Type I Models.\par

\tx Part of our plans for future research are aimed at path-integral
quantization of gravity using non-standard phase space variables.
In general this will be difficult to achieve since Eqs. (2.6) do not
not necessarily admit an action which can be calculated for different
histories.  In the present article we will do a preliminary calculation
in order to show how such a quantization might be expected to come out
using a diagonal Bianchi type I cosmology where the results of standard
path-integral quantization are known and the Hilbert action calculated
for the model is a valid action for its equations of motion.

\tx The form of the metric is$^9$
$$ds^2 = -N^2 d\alpha^2 + e^{2\alpha} e^{2\beta}_{ij} dx^i dx^j, \eqno (3.1)
$$
where $\beta_{ij} = \rm{diag} (\beta_{+} + \sqrt{3} \beta_{-}, \beta_{+} -
\sqrt{3} \beta_{-} , -2\beta_{+} )$, and ${{1}\over {6}}\ln g$ $={{1}\over
{6}} \ln [\rm{det}(g_{ij})]$ $= \alpha$ is taken as an internal time.  The
action can be reduced to the form
$$ I = \int [p_{+} \dot \beta_{+} + p_{-} \dot \beta_{-} - p_{\alpha} -
\tilde N (-p_{\alpha}^2 + p_{+}^2 + p_{-}^2 )]d\alpha, \eqno (3.2)$$
where $\cdot \equiv d/d\alpha$ and $\tilde N$ is a normalized lapse
function such as that used by Berger and Voegli$^{10}$.
Variation of this action with
respect to $p_{\pm}$, $\beta_{\pm}$, $p_{\alpha}$ and $\tilde N$ along
with the equation $\dot p_{\alpha} = 0$ give the full set of equations
for the model.  Our non-standard variable set will be $p_{\pm}$,
$\beta_{\pm}^{(0)}$, $K$, and $C = -p_{\alpha}^2 + p_{+}^2 + p_{-}^2$, where
the $\beta_{\pm}^{(0)}$ are constants of motion that are the initial
values of $\beta_{\pm}$,{\it i.e.\/}
$\beta_{\pm} (\alpha = 0)$, $p_{\pm}$ are
unchanged, $C$ is the non-linear combination given above, and $K$ is
an extension  of the phase space that we need to obtain the full equations
of motion.  In terms of these variables an action that gives the
correct classical equations of motion is
$$ I = \int [p_{+} \dot \beta_{+}^{(0)} + p_{-} \dot \beta_{-}^{(0)} +
C\dot K - \tilde N C]d\alpha. \eqno (3.3)$$
Varying $I$ with respect to $p_{\pm}$, $\beta_{\pm}^{(0)}$, $C$, $K$
and $\tilde N$ we find
$$ \dot p_{\pm} = \dot \beta_{\pm}^{(0)} = 0 , \eqno (3.4a)$$
$$ \dot C = \dot K = 0, \qquad C = 0. \eqno (3.4b)$$
The classical map between these variables and $\beta_{\pm}$, $p_{\alpha}$
($p_{\pm}$ being unchanged) gives the usual relations
$$p_{\pm} = p_{\pm}^{(0)} = \rm{const.}, \qquad p_{\alpha} = \sqrt{
p_{+}^{(0)2} + p_{-}^{(0)2}}, \qquad \beta_{\pm} = \beta_{\pm}^{(0)} +
{{p_{\pm}^{(0)} \alpha}\over {\sqrt{p_{+}^{(0)2} + p_{-}^{(0)2}}}}.
\eqno (3.5)$$

\tx The BRST path integral quantization of this system is relatively easy
to carry out.  What we would like to do is use a different coordinate
system on the space of paths than that generated by skeletonization.
That is, we would like to expand the possible paths in Fourier series,
something that is rarely done except for the harmonic oscillator, but,
of course, is possible for any continuous path.

\tx For BRST quantization we have to extend the phase space to a set of both
normal and anticommuting (ghost) variables.  As is usual in phase space
path-integral quantization, one must treat ``momentum'' and ``coordinate''
variables differently, and the decision about how to do this is somewhat of
an art.  In the Fourier-series formulation one way to handle this is
to decide on the type of Fourier series to be used for each variable.
We will not discuss this problem in detail here, but it will become obvious
in the type of series we choose.  The phase space is extended to include
$\Pi$, a momentum conjugate to $\tilde N$ and $\rho$, $\bar \rho$, $c$,
$\bar c$, four anticommuting functions of $\alpha$, which is all we need
since there is only one constraint, $C = 0$ and one gauge fixing.  We will
take the proper time gauge, so the gauge function $\chi (p_{\pm}, \beta_{\pm}
^{(0)}, K , C, \tilde N, \alpha) \equiv \dot {\tilde N}$ is zero, so the
gauge-fixing potential $\Phi$ is simply $\bar \rho \tilde N$, while the
BRST charge $\Omega$ has the form $\Omega = cC + \rho \Pi$ $^{11}$.  With all
of
these choices the Batalin-Fradkin-Vilkovisky$^{12}$ action for the system
described by (3.3),
$$I_B = \int [p_{+} \dot \beta_{+}^{(0)} +p_{-} \dot \beta_{-}^{(0)} +
C\dot K - \tilde N C + \bar \rho \dot c + \bar c \dot \rho + \Pi
\dot {\tilde N} - \{\Phi , \Omega\} ]d\alpha, \eqno (3.6)$$
becomes
$$ I_B = \int [p_{+} \dot \beta_{+}^{(0)} + p_{-} \dot \beta_{-}^{(0)} +
C \dot K - \tilde N C + \bar \rho \dot c + \bar c \dot \rho + \Pi
\dot {\tilde N} - \bar \rho \rho]d\alpha. \eqno (3.7)$$
The Fourier series we need for each of the variables which describe paths
between some value of the variables at $\alpha =0$ and other values at
$\alpha = T$ are
$$ \Pi = \sum_{n=1}^{\infty} \Pi_n \sin \left ( {{n\pi \alpha}\over {T}}
\right) ; \qquad \tilde N = N_0 + \sum_{n=1}^{\infty} N_n \cos \left (
{{n\pi \alpha}\over {T}} \right ); \eqno (3.8a)$$
$$ c = \sum_{n=1}^{\infty} c_n \sin \left( {{n\pi \alpha}\over {T}} \right );
\qquad \rho = \rho_0 + \sum_{n=1}^{\infty} \rho_n \cos \left ( {{n\pi \alpha
}\over {T}} \right ); \eqno (3.8b)$$
$$ \bar c = \sum_{n=1}^{\infty} \bar c_n \sin \left ({{n\pi \alpha}\over
{T}} \right ); \qquad \bar \rho = \bar \rho_0 + \sum _{n=1}^{\infty} \bar
\rho_n \cos \left ( {{n\pi \alpha}\over {T}} \right ); \eqno (3.8c)$$
$$ \beta_{\pm}^{(0)} = \beta_{\pm c}^{(0)} (\alpha) + \sum_{n=1}^{\infty}
\beta_{\pm}^{(0)n} \sin  \left ({{n\pi \alpha}\over {T}} \right ) ; \qquad
K = K_c (\alpha) + \sum_{n=1}^{\infty}
K^{(n)} \sin \left ( {{n\pi \alpha}\over {T}}
\right ) ; \eqno (3.8d)$$
$$ p_{\pm} = p_{\pm}^{(0)} + \sum_{n=1}^{\infty} p_{\pm}^{(n)} \cos \left
( {{n\pi \alpha}\over {T}} \right ) ; \qquad C = C^{(0)} + \sum_{n=1}
^{\infty} C^{(n)} \cos \left ( {{n\pi \alpha}\over {T}} \right ), \eqno
(3.8e)$$
where the Fourier coefficients of the anticommuting
variables are anticommuting
numbers, and, in principle functions with subscript $c$ are the classical
solutions for those variables.  Since the variables that appear in (3.8d)
have classical solutions that are constant, $\beta_{\pm c}^{(0)}$ and $K_c$
would be constants.  These classical solutions give unphysical results, so
we will use a linear solution for each of them connecting $\beta_{\pm 0}
^{(0)}$, $K_0$ at $\alpha =0$ with $\beta_{\pm 1}^{(0)}$, $K_1$ at
$\alpha = T$, or
$$\beta_{\pm c}^{(0)} = {{1}\over {T}} (\beta_{\pm 1}^{(0)} - \beta_{\pm 0}
^{(0)} ) \alpha + \beta_{\pm 0}^{(0)}; \qquad K_c = {{1}\over {T}}(K_1 - K_0)
\alpha + K_0. \eqno (3.9)$$

\tx The ghost part of the action $I_{gh} = \int_0^T [\bar \rho \dot c +
\bar c \dot \rho - \bar \rho \rho ]d\alpha$ becomes
$$ I_{gh} = \sum_{n=1}^{\infty} [\bar \rho_n c_n {{n\pi}\over {2}} -
\bar c_n \rho_n {{n\pi}\over {2}} - \bar \rho_n \rho_n T/2] -\bar \rho_0
\rho_0 T. \eqno
(3.10)$$
The ghost part of the propagator $\int \prod_n dc_n \prod_n d\bar c_n \prod
_n d\rho_n \prod_n d\bar \rho_n d\rho_0 d\bar \rho_0 e^{iI_{gh}}$ can be
shown, using Berezin integration (with the normalization $\int \theta d\theta
= 1$)$^{13}$, to be

$T[-i\prod_n ({{n^2 \pi^2}\over {4}} )]$.  The rest of the
action, $I_A$ is
$$ I_A = \int_0^T [ \{p_{+}^{(0)} + \sum_{n=1}^{\infty} \cos \left
({{n\pi \alpha}\over {T}} \right ) \} \{ ({{1}\over {T}})(\beta_{+1}^{(0)}
- \beta_{+0}^{(0)} ) + \sum_{n=1}^{\infty} {{n\pi}\over {T}} \beta_{+}^{(n)}
\cos \left( {{n\pi \alpha}\over {T}} \right ) \} +$$
$$ +\{p_{-}^{(0)} + \sum_{n=1}^{\infty} p_{-}^{(n)} \cos \left(
{{n\pi \alpha}\over {T}} \right ) \} \{ ({{1}\over {T}})(\beta_{-1}^{(0)} -
\beta_{-0}^{(0)} ) + \sum_{n=1}^{\infty} {{n\pi}\over {T}} \beta_{-}^{(n)}
\cos \left ( {{n\pi \alpha}\over {T}} \right ) \}+$$
$$+\{ C^{(0)} + \sum_{n=1}^{\infty} C^{(n)} \cos \left ( {{n\pi \alpha}\over
{T}} \right ) \} \{ ({{1}\over {T}}) (K_1 - K_0) + \sum_{n=1}^{\infty}
{{n\pi}\over {T}} K^{(n)} \cos \left
( {{n\pi \alpha}\over {T}} \right ) \} -$$
$$- \sum_{n=1}^{\infty} \Pi_n \sin \left ( {{n\pi \alpha}\over {T}} \right)
\sum_{m=1}^{\infty} N_m {{m\pi}\over {T}} \sin \left ( {{m\pi \alpha}\over
{T}} \right ) -$$
$$ - \{ N_0 + \sum_{n=1}^{\infty} N_n \cos \left ( {{n\pi \alpha}\over
{T}} \right ) \} \{ C^{(0)} + \sum_{n=1}^{\infty} C^{(n)} \cos \left (
{{n\pi \alpha}\over {T}} \right ) \}  ] d\alpha $$
$$= p_{+}^{(0)} (\beta_{+1}^{(0)} - \beta_{+0}^{(0)} ) + \sum p_{+}^{(n)}
\beta_{+}^{(n)} {{n\pi}\over {2}} + p_{-}^{(0)} (\beta_{-1}^{(0)} -
\beta_{-0}^{(0)} ) + $$
$$ + \sum_{n=1}^{\infty} p_{-}^{(n)} \beta_{-}^{(n)} {{n\pi}\over {2}} +
C^{(0)} (K_1 - K_0) + \sum_{n=1}^{\infty} C^{(n)} K^{(n)} {{n\pi}\over {2}}+
$$
$$+ \sum_{n=1}^{\infty}	\Pi_n N_n {{n\pi}\over {2}} - N^{(0)} C^{(0)} T +
\sum_{n=1}^{\infty} N_n C^{(n)} {{n\pi}\over {2}}. \eqno (3.11)$$
Integrating $e^{iI_A}$ over $\Pi_n$ (we will always use the Liouville
measure, {\it e.g.\/} for $\Pi_n$, $d\Pi_n /2\pi$) from $-\infty$ to
$+\infty$ gives an infinite product of
delta functions of the form $\delta ({{n\pi}\over {2}} N_n)$ which
when integrated over the $N_n$ removes the last term of (3.11) at the cost
of a term $\prod _n (1/n\pi^2)$ as an overall factor. Similar integrations
over $C^{(n)}$, $p_{\pm}^{(n)}$ and subsequently $K^{(n)}$ and $\beta_{\pm}
^{(n)}$ remove all of the summations at the cost of three more $\prod_n (
1/n\pi^2)$ factors.  The integration over $N^{(0)}$ gives $\delta (C^{(0)} T)
= (1/T) \delta (C^{(0)})$ which removes the factor of $T$ in the ghost
integration and also gets rid of the $C^{(0)} (K_1 - K_0)$ term.  The
final form of the propagator is
$$<\beta_{\pm 1}^{(0)}, K_1, T \mid \beta_{\pm 0}^{(0)}, K_0, 0> =
-i\prod_n \left ({{n^2 \pi^2}\over {4}} \right ) \left \{ \left (
\prod_{\ell} {{1}\over {\ell \pi^2}} \right )^4 \right \} \times$$
$$\times \int_{-\infty}
^{\infty} {{dp_{+}^{(0)}}\over {2\pi}} {{dp_{-}^{(0)}}\over {2\pi}} e^
{ip_{+}^{(0)} (\beta_{+1}^{(0)} - \beta_{+0}^{(0)} )} e^{ip_{-}^{(0)} (
\beta_{-1}^{(0)} - \beta_{-0}^{(0)} )}$$
$$ = {\cal N} \delta (\beta_{+1}^{(0)} - \beta_{+0}^{(0)} ) \delta (
\beta_{-1}^{(0)} - \beta_{-0}^{(0)}). \eqno (3.12)$$
The constant normalization ${\cal N}$ can be treated as one over the
Jacobian of the transformation from skeletonization coordinates to
Fourier-series coordinates on the space of paths as Feynman does for the
case of the harmonic oscillator.

\tx Notice that the Fourier series coordinates are simpler to use than
skeletonization for linear equations of motion because the coefficients are
nicely grouped to give $\delta$-functions that kill terms on further
integration.  Unfortunately, this simplicity disappears for more
complicated motions unless one can calculate explicitly the Fourier series
for the motion and integrate easily over the resulting coefficients.

\tx The propagator above can be compared to the Green function for the
Schr\"odinger quantization of the system, and they agree, since the
solution for $\Psi (\beta_{\pm}^{(0)}, K, \alpha ) = \Psi (\beta_{\pm}
^{(0)})$, that is, an arbitrary function of $\beta_{\pm}^{(0)}$ that is
independent of time.  This wave function is perhaps the ultimate in
frozen dynamics, but, of course, this fact means little until one
knows how to map such a wave function into a true Hilbert space function.
In principle one must develop a kernel such as those mentioned in Sec. 2 to
map $\Psi (\beta_{\pm}^{(0)})$ to $\psi (\beta_{\pm} , \alpha)$, but we will
leave this to future work.  Here we will appeal to an argument similar
that used in our Bianchi-V work, where we argued that the $\partial /
\partial t$ that appears in the Schr\"odinger equation  is a partial
derivative that implies holding certain variables constant$^1$.  However,
since the variables $\beta_{\pm}^{(0)}$ depend on $\alpha$ implicitly,
the function $\Psi$ can depend on $\alpha$ through this dependence.  That
is, since $\beta_{\pm}^{(0)} = \beta_{\pm} - (p_{\pm}^{(0)}/ \sqrt {p_{+}
^{(0)2} + p_{-}^{(0)2}})\alpha$, $\Psi$ can be written as
$$\Psi = \Psi \left ( \beta_{\pm} - {{p_{\pm}^{(0)} \alpha}\over
{\sqrt {p_{+}^{(0)2} + p_{-}^{(0)2}}}} \right). \eqno (3.13)$$
In fact, the product of eigenstates of $p_{\pm}$ with eigenvalue
$p_{\pm}^{(0)}$ becomes
$$ \Psi = e^{i(p_{+}^{(0)} \beta_{+} + p_{-}^{(0)} \beta_{-} - \sqrt
{p_{+}^{(0)2} + p_{-}^{(0)2}} \alpha)}, \eqno (3.14)$$
which is the solution found by Charlie in his original study of quantum
cosmology$^9$.

\tx As we mentioned in the Introduction, we have not attempted to carry this
path integral formulation too far.  We have only tried to give the flavor
of the quantization of the gravitational field using non-standard phase
space variables by presenting this simple model.  Path integral quantization
in non-standard phase space variables still faces many problems, especially
for systems that do not come from a variational principle, where even the
definition of the process will require new ideas, and it may even be
impossible to achieve a consistent theory.

\ha References\par

\tx 1. S. Hojman, D. N\'u\~nez, and M. Ryan, {\it Phys. Rev. D\/} {\bf 45},
3523 (1992).

2. V. Bargmann, {\it Comm. Pure and App. Math.\/} {\bf 14}, 2960 (1961).

3. F. Dyson, {\it Am. J. Phys.\/} {\bf 58}, 209 (1990).

4. S. Hojman and L. Shepley, {\it J. Math. Phys. \/} {\bf 32}, 142 (1991)

5. S. Hojman, in press

6. See, for example, D. Simms and N. Woodhouse, {\it Lectures on
Geometric\/}

\hskip 3 em {\it Quantization\/}
(Springer, Berlin, 1977);

\hskip 3 em N. Woodhouse, {\it Geometric Quantization\/}
(Oxford U. P., Oxford, 1981).

7. A. Ashtekar {\it et al.\/},
{\it New Perspectives in Canonical Gravity\/}
(Bibliopolis,

\hskip 3 em Naples, 1988).

8. V. Fock, {\it Z. Phys.\/} {\bf 49}, 339 (1928).

9. C. Misner, {\it Phys. Rev.\/} {\bf 186}, 1319 (1969).

10. B. Berger and C. Voegli, {\it Phys. Rev. D\/} {\bf 32}, 2477 (1985).

11. See J. Guven and M. Ryan, {\it Phys. Rev. D\/} {\bf 45}, 3559
(1992),
and

\hskip 3 em references therein.

12. L. Fradkin and G. Vilkovisky, CERN Report TH 2332 (1977);

\hskip 3 em I. Batalin and G. Vilkovisky, {\it Phys. Lett.\/} {\bf 69B},
309 (1977).

13. F. Berezin, {\it The Method of Second Quantization\/} (Academic Press,

\hskip 3 em New York, 1966).
\end